\documentclass[twocolumn,showpacs,preprintnumbers,amsmath,amssymb,prl]{revtex4}

\usepackage{graphicx}
\usepackage{dcolumn}
\usepackage{bm}
\usepackage{color}

\begin{document}

\title{Quantum Memory with Optically Trapped Atoms}

\author{Chih-Sung Chuu$^1$}
\thanks{These authors contributed equally to this work}
\thanks{chuu@physi.uni-heidelberg.de}
\author{Thorsten Strassel$^1$}
\thanks{These authors contributed equally to this work}
\thanks{strassel@physi.uni-heidelberg.de}
\author{Bo Zhao$^1$}
\author{Markus Koch$^1$}
\author{Yu-Ao Chen$^{1,2}$}
\author{Shuai Chen$^1$}
\author{Zhen-Sheng Yuan$^{1,2}$}
\author{J{\"o}rg Schmiedmayer$^{3}$}
\author{Jian-Wei Pan$^{1,2}$}
\affiliation{%
$^1$Physikalisches Institut, Universit\"at Heidelberg, D-69120 Heidelberg, Germany\\
$^2$Hefei National Laboratory for Physical Sciences at Microscale, Department of Modern Physics, \\
University of Science and Technology of China, Heifei 230026, China \\
$^3$Atominstitut der {\"O}sterreichischen Universit{\"a}ten, TU-Wien, A-1020 Vienna, Austria
}%

\date{\today}

\begin{abstract}
We report the experimental demonstration of a quantum memory for collective atomic states in a far-detuned optical dipole trap. Generation of the collective atomic state is heralded by the detection of a Raman scattered photon and accompanied by storage in the ensemble of atoms. The optical dipole trap provides confinement for the atoms during the quantum storage while retaining the atomic coherence. We probe the quantum storage by cross-correlation of the photon pair arising from the Raman scattering and the retrieval of the atomic state stored in the memory. Non-classical correlations are observed for storage times up to 60 $\mu$s.

\end{abstract}

\pacs{03.67.Hk, 37.10.Gh, 42.50.Dv}

\maketitle

A quantum memory, a storage device for quantum states, is requisite to a scalable quantum repeater \cite{QuantumRepeater} for the realization of long-distance quantum communication~\cite{QuantumCommunication}. In the quantum repeater protocol, the tranmission channel is divided into several segments with lengths comparable to the channel attenuation length. Entanglement is then generated and purified \cite{Purification} for short distances before being extended to a longer distance by entanglement swapping \cite{Swapping}. The entanglement creation, purification, and connection are probabilistic, thereby requiring the successfully entangled segment state to be stored in a quantum memory while waiting for the others to generate. Once the entanglement is distributed over the transmission channel, it can be used for quantum teleportation \cite{Teleportation} or cryptography \cite{Cryptography}. A quantum memory with long storage time is therefore crucial to achieve scalable quantum communication networks with a manageable time overhead.

Various schemes were proposed for implementing quantum repeaters \cite{DLCZ,Pan5,QR3} in which the scalability stems from the entanglement between a sent photon and the quantum state stored in the quantum memory. The quantum state is stored in a collective state of an atomic ensemble where a superposition between two ground states is shared among all the atoms. The key issue to a quantum memory is that the stored state, which could be later read out by converting into another photon, keeps its quantum correlation with the sent photon. This correlation also allows an arbitrary state to be written into the quantum memory by quantum teleportation.

Significant progress has been made toward realization of quantum repeaters in recent years. Non-classical correlation has been observed between Raman scattered photons and the consequent collective excitations in an atomic ensemble \cite{Kimble3,Lukin2,Harris1,Pan1,Vuletic1}. Number-state entanglement has also been generated between two ensembles of atoms \cite{Kuzmich2,Kimble5}. Most recently, quantum teleportation with a built-in quantum memory \cite{Pan2} and entanglement swapping \cite{Pan6} have been demonstrated. 

Despite these advances, the storage times of the quantum memories reported to date are primarily limited by inhomogeneous broadening of the ground state due to magnetic fields \cite{Kimble2}. As a result, the quadrupole magnetic field of the magneto-optical trap (MOT) used to confine the atoms is switched off during the storage. Nevertheless, the residual magnetic field still limits the lifetime (see below) of the quantum memory to only $\sim$~10~$\mu$s~\cite{Pan1,Pan2,Kimble2}. Moreover, without confinement the atoms simply diffuse out of interaction region ($\sim$~100~$\mu$m) after a few hundred microseconds. For a quantum memory in a MOT, the atomic diffusion imposes a strict limitation on the storage time, which in turn limits the maximum distance for quantum communication in practical applications.

On the contrary, a quantum memory confined in a far-detuned optical dipole trap \cite{Grimm} can simultaneously achieve long coherence and confinement as widely discussed in the recent literature \cite{Kimble3,Pan2,Vuletic1,Pan6,Kimble2}. In an optical trap, the large detuning efficiently suppresses photon scattering, which results in a non-dissipative confinement for the atoms without introducing magnetic fields. Furthermore, coherent manipulation of atomic internal states for seconds has been demonstrated in such traps~\cite{Chu1}. A quantum memory of this sort thus has a potential storage time of seconds. However, realization of a memory in an optical trap is extremely challenging and, to the best of our knowledge, has not been achieved.

In this Letter, we experimentally demonstrate a quantum memory in a focused-beam optical trap that has a long trap lifetime ($\tau_{trap}=$ 20~s) and a negligible photon scattering rate ($\Gamma =$ 4~s$^{-1}$). In addition to optical trapping, we also make use of the first-order magnetic-field-insensitive state (clock state). For the clock state, we choose the superposition state of the magnetic sublevels, $|c_{1} \rangle$ = $|5S_{1/2}, F = 1, m_{F} = -1\rangle$ and $|c_{2}\rangle$ = $|5S_{1/2} , F = 2, m_{F} = 1\rangle$, of $^{87}$Rb. These states experience the same first-order Zeeman shift at a magnetic field of $\sim$~3.23~G~\cite{clockstate}. Hence, an atomic ensemble in a superposition of this state pair is, to first-order, insensitive to the spatial inhomogeneity and temporal fluctuation of the magnetic field \cite{Cornell2,Hansch1}. Our quantum memory is therefore robust against the principal decoherence process experienced in previous experiments and provides an additional confinement during quantum storage.

\begin{figure}
\includegraphics[width=0.9\linewidth]{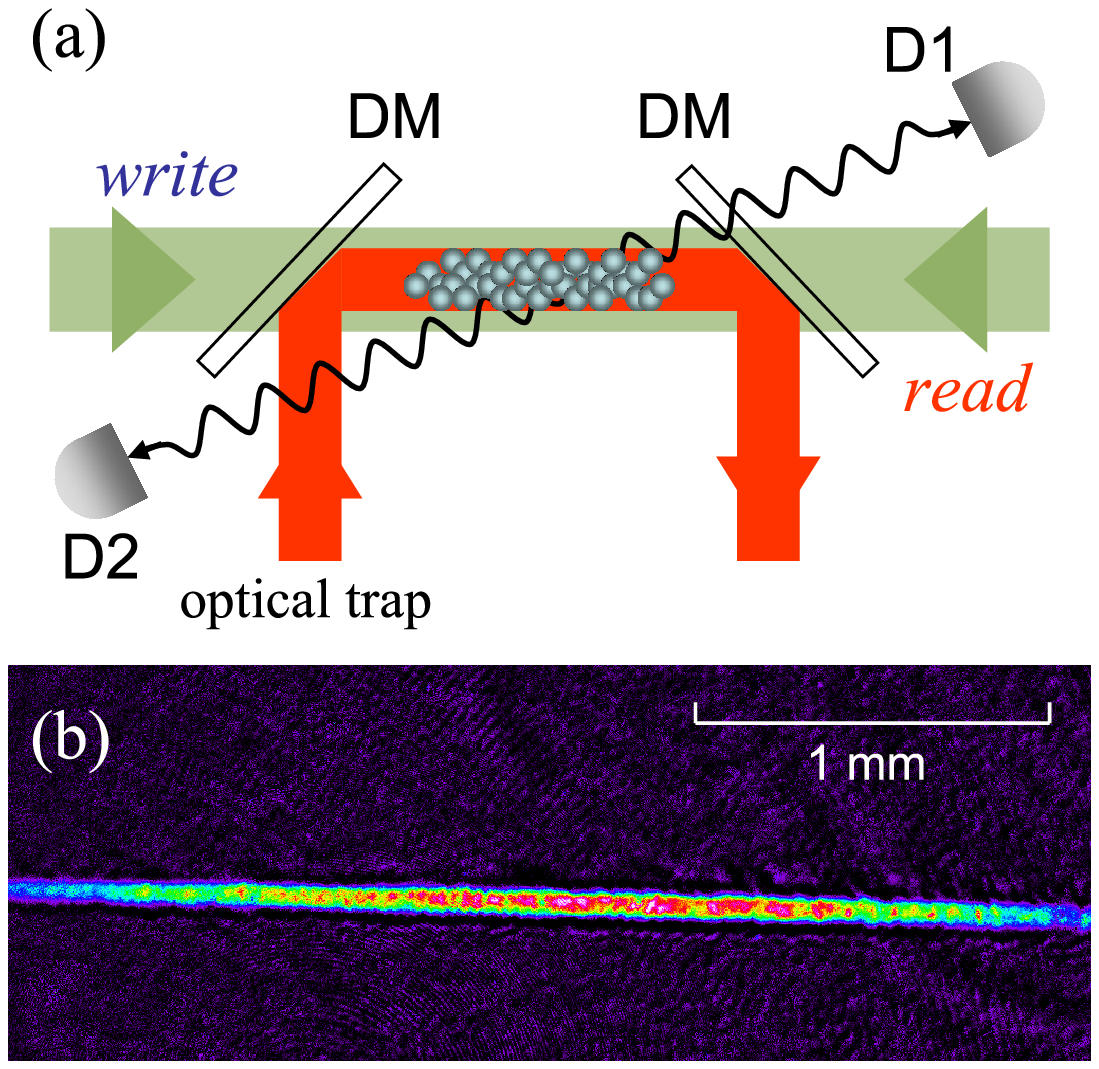}
\caption{\label{fig:1} (color online) A schematic of our apparatus. (a) The atomic ensemble is confined in an optical trap formed by a red-detuned, focused laser beam. The optical-trapping beam is overlapped with the \textit{write} and \textit{read} beams (counterpropagating to each other) on the dichroic mirrors (DM). Stokes and anti-Stokes photons are detected by single-photon detectors D1 and D2, respectively, at an angle of $\sim$~2$^{\circ}$ with respect to the optical-trapping beam. (b) An absorption image of the optically trapped atoms along the radial direction after a time-of-flight of 0.3 ms.}
\end{figure}

\begin{figure}
\includegraphics[width=0.9\linewidth]{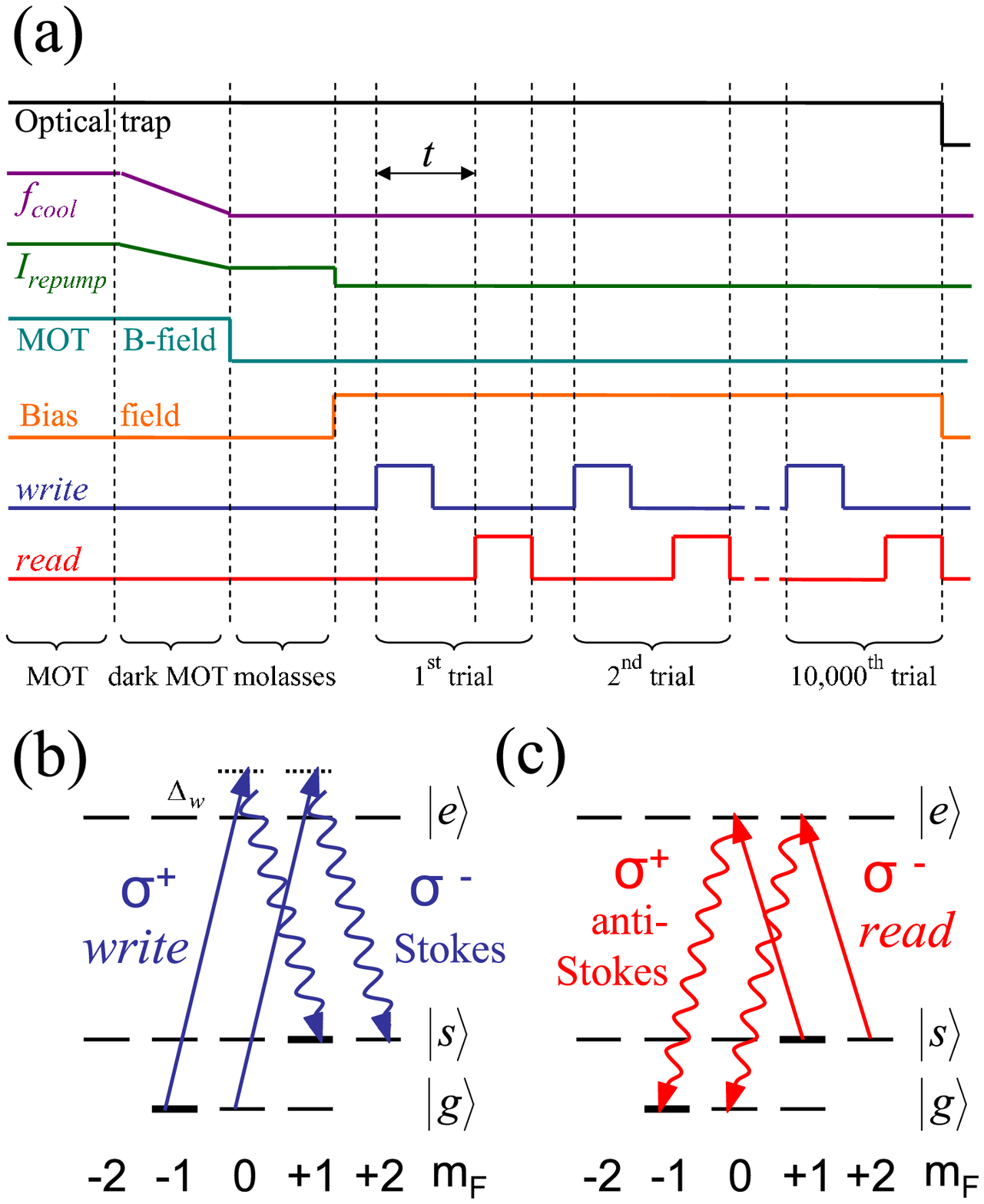}
\caption{\label{fig:2} (color online) Time sequence and relevant atomic transitions of the experiment. (a) After the MOT is switched off, atoms are transfered into an optical trap. Subsequently, 10,000 alternating \textit{write} and \textit{read} pulses, with pulse lengths of 100 ns and 500 ns respectively and a controllable time delay of $t$, illuminate the atomic ensemble. Before the first \textit{write} pulse, a \textit{read} pulse (not shown) is applied to ensure that no atoms are in the metastable state $|s\rangle$. (b) and (c) illustrate the relevant atomic levels involved in the \textit{write} and \textit{read} processes respectively, where $|g\rangle=|5S_{1/2},F=1\rangle$, $|s\rangle=|5S_{1/2},F=2\rangle$, and $|e\rangle=|5P_{1/2},F=2\rangle$. Zeeman splitting is not shown in this figure.}
\end{figure}

A schematic of our apparatus is shown in Fig.~\ref{fig:1} (a). The experiment begins with a standard MOT. During 2~s of loading, $5 \times 10^6$ $^{87}$Rb atoms, with a temperature of about 100~$\mu$K, are loaded from the background vapor. We then change the configuration of the MOT to a temporal dark MOT, followed by molasses cooling, to maximize the transfer of atoms into the optical trap. In the dark MOT phase, the frequency of the cooling light, $f_{cool}$, is shifted to the red of the $|5S_{1/2},F=2\rangle \rightarrow |5P_{3/2},F=~3\rangle$ transition by 35~MHz for a duration of 145 ms, while the repumping intensity $I_{repump}$ is ramped down by a factor of 200. After the dark MOT, the MOT quadrupole field is switched off and molasses cooling is applied for 5 ms, resulting in a peak atomic density of $3\times 10^{10}$ cm$^{-3}$ and a temperature of about 20~$\mu$K. Subsequently, a magnetic bias field at $\sim$ 3.23 G is switched on along the longitudinal direction of the atomic cloud. The atoms are then optically pumped to the $|g\rangle$ = $|5S_{1/2},F=1 \rangle$ hyperfine state by shuttering off the repumping light 3~ms before the cooling light is extinguished. The time sequence of our experiment is summarized in Fig.~\ref{fig:2}~(a).

The optical trap is formed by a tightly focused laser beam at $\lambda=1030$ nm with a $1/e^{2}$ radius of 36 $\mu$m. The beam is left on during the experimental cycles at 7~W, leading to radial and axial trapping frequencies of $\omega_r = 2 \pi \times$2~kHz and $\omega_z = 2 \pi \times$10 Hz, respectively, and a trap depth of $k_{B} \times 500$ $\mu$K, with $k_{B}$ denoting the Boltzmann's constant. For typical operating conditions, $2 \times 10^{5}$ atoms are transfered into the optical trap (untrapped atoms are allowed to free fall for 30 ms). The temperature of the atoms after the transfer increases to 45~$\mu$K, possibly due to the heating associated with the optical pumping. Together with the measured radial and axial rms radius of 5.5 $\mu$m and 0.85~mm, respectively, we derive a peak atomic density of $10^{12}$~cm$^{-3}$ in the optical trap. Fig.~\ref{fig:1}~(b) shows an absorption image of the optically trapped atoms after a time-of-flight of 0.3~ms.

With all atoms prepared in the ground state $|g\rangle$, a weak \textit{write} pulse illuminates the atomic ensemble for a duration of 100~ns. Fig.~\ref{fig:2} (b) illustrates the relevant atomic transitions in the experiment. The right-circularly ($\sigma^{+}$) polarized \textit{write} beam, co-propagating with the optical-trapping beam, is blue-detuned $\Delta_w =$ 100 MHz from the $|g\rangle \rightarrow |e\rangle$ = $|5P_{1/2},F=2\rangle$ transition with a $1/e^{2}$ radius of 250 $\mu$m. Each \textit{write} pulse contains approximately $10^5$ photons in the region of the atoms. As a result, there is an extremely small probability ($p_S \ll 1$) of inducing spontaneous Raman transition to the metastable state $|s\rangle$ = $|5S_{1/2},F=2\rangle$ and detecting a Stokes photon with left-circularly ($\sigma^{-}$) polarization by a single-photon detector D1. The collection mode of detector D1, with a $1/e^{2}$ radius of 75~$\mu$m, is tilted at an angle of $\sim$ 2$^{\circ}$ with respect to the \textit{write} beam for frequency filtering. 

Prior to the detection of a Stokes photon, the joint state of the atomic collective mode and the Stokes mode is described by $|\phi \rangle \propto |0_{a} \rangle |0_{S} \rangle + e^{i \beta} \sqrt{p^{}_{S}} |1_{a} \rangle |1_{S} \rangle + O(p^{}_{S})$~\cite{DLCZ}, where $|i_{a} \rangle$ and $|i_{S} \rangle$ denote $i$ quanta of excitations in the atomic collective and Stokes modes, respectively, $\beta$ is a phase related to the propagation of the Stokes photon \cite{Kimble2}, and $O(p^{}_{S})$ represents terms with higher excitations. Conditional upon a click at detector D1, the Stokes field is projected to $|1_S \rangle$ and a collective atomic superposition state,
\begin{equation}
\label{collective}
|1_{a}\rangle = \frac{1}{\sqrt{N}} \sum^{N}_{j=1} e^{-i(\textbf{k}_{s}-\textbf{k}_{w})\cdot\textbf{r}_{j}} |g\rangle_{1} \cdots |s\rangle_{j} \cdots |g \rangle_{N}, 
\end{equation}
is created and stored in the ensemble. Here, $N$ is the atom number, $\textbf{k}_{S}$ and $\textbf{k}_{w}$ are the wave vectors of the Stokes and \textit{write} photons, respectively, and $\textbf{r}_{j}$ is the position of the $j$th atom when it scatters a Stokes photon. 

After a controllable time delay (storage time) $t$, the collective excitation in the atomic ensemble is converted into a photon in the anti-Stokes mode, $|1_{AS} \rangle$. In the experiment, only photons with $\sigma ^{+}$ polarization are detected by detector D2, for which the collection mode is matched to detector D1 with a coupling efficiency of 80$\%$. The coherent conversion is accomplished by illuminating the atoms with a strong, $\sigma ^{-}$ polarized $\textit{read}$ beam that is resonant with the $|s\rangle \rightarrow |e\rangle$ transition and mode-matched to the $\textit{write}$ beam in a counter-propagating configuration. This results in a non-classical pair of Stokes and anti-Stokes photons. The number correlation of the photon pair is thus a useful probe for the storage in the quantum memory, as various sources of decoherence could cause degradation.

We explore the number correlation of the Stokes and anti-Stokes photons by measuring the normalized cross-correlation function, or the quantum mechanical degree of second-order coherence, of the two fields~\cite{Kimble3}, $g^{}_{S,AS} = p^{}_{S,AS}/p^{}_{S}p^{}_{AS}$, where $p^{}_{S,AS}$ is the joint probability of detecting one photon by both detectors within the same experimental trail, and $p^{}_S$ (or $p^{}_{AS}$) is the probability of detecting a Stokes (or anti-Stokes) photon individually. For a non-classical photon pair, $g^{}_{S,AS}$ violates the classical constraint set by the Cauchy-Schwarz inequality \cite{Clauser}, $g^{2}_{S,AS}\leq g^{}_{S,S}g^{}_{AS,AS}$, where $g^{}_{S,S}$ and $g^{}_{AS,AS}$ are the normalized auto-correlation functions \cite{autocorrelation}. In our experiment, the anti-Stokes and Stokes fields are both in thermal states, $g^{}_{S,S}=g^{}_{AS,AS}=2$ \cite{Scully}, and therefore measuring $g^{}_{S,AS} > 2$ is an indication of non-classical correlation.

\begin{figure}
\includegraphics[width=0.8\linewidth]{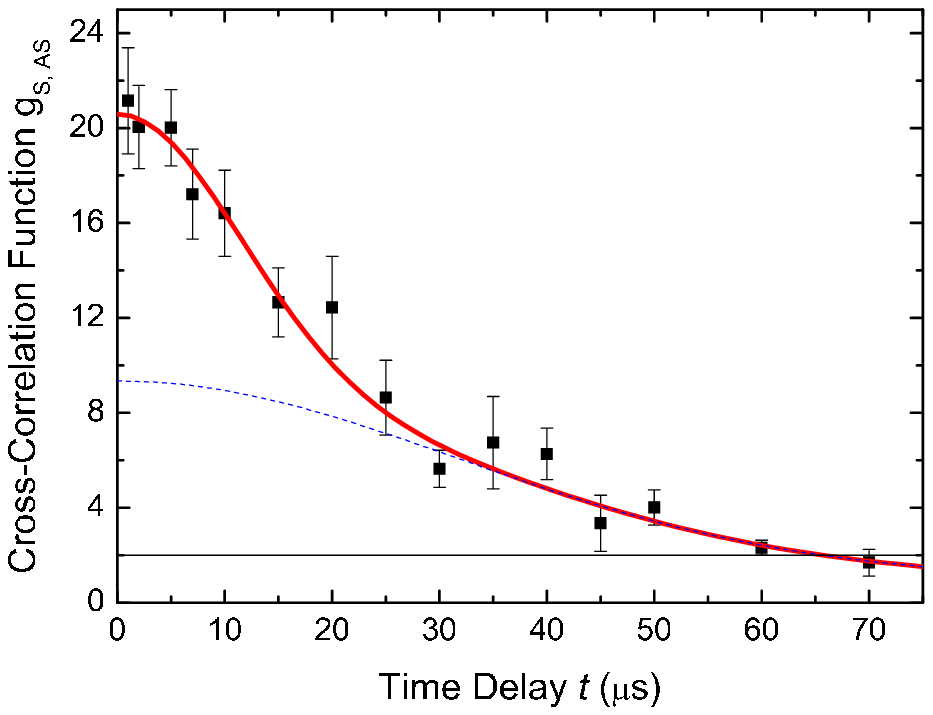}
\caption{\label{fig:3} (color online) Normalized cross-correlation of the Stokes and anti-Stokes fields as a function of the time delays between the \textit{write} and \textit{read} pulses. Non-classical correlation is observed for storage times up to 60 $\mu$s. The curve is a Gaussian fit with two time constants, $\tau_{nc} = 16(3)$~$\mu$s and $\tau_c = 45(5)$~$\mu$s, and its asymptotic value of 1 as time delay $t \gg \tau_c$ corresponds to the probabilistic coincidence events, $p^{}_{S,AS}=p^{}_S p^{}_{AS}$. The dotted line is a decay curve with only one time constant $\tau_c$, assuming the absence of atoms in the non-clock state. The horizontal line at $g^{}_{S,AS}=2$ indicates the onset of quantum correlation. The error bars indicate the statistical errors.}
\end{figure}

Fig.~\ref{fig:3} illustrates our main result where the normalized cross-correlation function is measured for various time delays between the \textit{write} and \textit{read} pulses. Violations of the Cauchy-Schwarz inequality are observed for delays up to 60 $\mu$s. The observed violations correspond to the temporal storage of the non-classical correlation between the Stokes and anti-Stokes photons as well as the atomic collective state in the quantum memory. The measured $g^{}_{S,AS}$ function follows a Gaussian decay with two time constants. The Gaussian dependence is a consequence of the inhomogeneous phase broadening of the collective state due to the residual magnetic field \cite{Kimble2} and the Maxwellian velocity distribution of the atoms~\cite{Bo}. The two time scales result from different Zeeman components of the atomic ensemble. During the \textit{write} process, both the $|c_{1}\rangle \rightarrow |e\rangle \rightarrow |c_{2}\rangle$ (clock) and $|F=~1,m_{F}=~0~\rangle \rightarrow |e\rangle \rightarrow |F=2,m_{F}=2\rangle$ (non-clock) transitions contribute to the detection of Stokes photons with $\sigma^{-}$ polarization. Likewise, the reversed transitions are involved in the detection of anti-Stokes photons with $\sigma^{+}$ polarization during the \textit{read} process.

The time dependence of the normalized cross-correlation function thus can be described by $g^{}_{S,AS}(t)=1+{\rm{A}}_{nc}e^{-(t/\tau_{nc})^2}+{\rm{A}}_{c}e^{-(t/\tau_{c})^2}$ \cite{Pan1}, where $\tau_c$ and $\tau_{nc}$ are the coherence times, or lifetimes, of the quantum memory, and the coefficients ${\rm{A}}_c$ and ${\rm{A}}_{nc}$ depend on the relative strengths of the clock and non-clock transitions as well as the initial atomic population in the Zeeman sublevels of the $F = 1$ manifold. Two lifetimes obtained from the fit to the measurement are $\tau_{nc}^{\rm{exp}} = 16(3)$~$\mu$s and $\tau_c^{\rm{exp}} = 45(5)$~$\mu$s. The fast decay in the short time scale is mainly due to the decoherence induced by the residual magnetic field for the atoms in the non-clock state. From $\tau_{nc}^{\rm{exp}}$, the inhomogeneity of the magnetic field is estimated to be approximately 10~mG, which is comparable to what has been observed in previous experiments \cite{Pan1,Pan2,Kimble2}. The slow decay in the long time scale, on the the hand, is associated with the atoms in the clock state. The corresponding decay time $\tau_{c}^{\rm{exp}}$ is well beyond the limit $\tau_{nc}^{\rm{exp}}$ imposed by the residual magnetic field, which indicates the robustness of the clock state.

The thermal motion of the atoms after the \textit{write} process accounts for the remaining decoherence in our experiment \cite{AtomicMotion}. The atomic motion induces dephasing in the collective state through the position-dependent phase factors in Eq.~\ref{collective}. During the quantum storage, an atom traveling with a velocity of $\textit{\textbf{v}}$ leads to an additional phase of $(\textbf{k}_{s}-\textbf{k}_{w})\cdot\delta \textbf{r}$ with $\delta \textbf{r} = \textbf{\textit{v}}t$. The consequent dephasing results in a decay in the normalized cross-correlation function with a time scale of $\tau_{c}^{\rm{th}} \sim 1/(v |\Delta \textbf{k}|)$, where $v$ is the velocity of the atom along the direction of $\Delta \textbf{k} = \textbf{k}_{s}-\textbf{k}_{w}$. In our experiment, $v$ and $|\Delta \textbf{k}|$ are estimated to be 66~mm/s and 290~mm$^{-1}$, respectively, and a lifetime of $\tau_c^{\rm{th}} \cong 53$~$\mu$s is inferred. The agreement between $\tau_{c}^{\rm{th}}$ and $\tau_{c}^{\rm{exp}}$ thus implies the preservation of coherence in the optical trap during the quantum storage. The discrepancy is likely due to the misalignment of the magnetic bias field.

In conclusion, we have realized a quantum memory with optically trapped atoms for collective atomic states. Non-classical correlations of the photon pair, arising from the spontaneous Raman scattering and the retrieval of the collective atomic state, are observed for storage times up to 60~$\mu$s. The measured lifetime of the quantum memory associated with the atoms in the clock state is beyond the limit imposed by the residual magnetic field, which shows the robustness of the clock state. Together with the non-dissipative confinement provided by the optical trap, the quantum memory has a potential storage time of seconds. The observed storage time is currently limited by the thermal motion of the atoms. With an atomic ensemble at a lower temperature, a longer storage time could be achieved. For example, a temperature of submicro-Kelvin could be obtained by employing evaporative cooling in a crossed optical trap \cite{Chapman1}, which will greatly reduce the atomic motion and thus extend the storage time beyond milliseconds. Alternatively, one could confine the atoms in an optical lattice and expect a storage time of seconds. We also note that the inhomogeneous light shift in the red-detuned optical trap could be improved by confining the atoms in a blue-detuned ``box'' trap \cite{OpticalBox}. Lastly, our experiment utilizes the atoms in the clock state as well as the non-clock state. One could also realize a quantum memory composed of atoms purely in the clock state, for instance, by optically pumping all atoms into the state $|c_{1} \rangle$ before each experimental trial. 

This work was supported by the Alexander von Humboldt Foundation, the MCE Grants, the DFG, the CAS, the NFRP, and the KAS.

\bibliography{dipoletrap}

\begin{thebibliography}{31}
\bibitem[{clo()}]{clockstate}
\bibinfo{note}{The differential Zeeman shift can be described by $\Delta \nu =
  \Delta \nu_{\rm{min}} + \beta (B - B_0)^2$, with $\Delta \nu_{\rm{min}}$ and
  $\beta$ given in ref. \cite{Cornell2}}.
\bibitem[{aut()}]{autocorrelation}
\bibinfo{note}{To measure $g^{}_{S,S}$ (or $g^{}_{AS,AS}$), one would direct
  the Stokes (or anti-Stokes) field onto a 50/50 beamsplitter with a
  single-photon detector placed behind each output port, then $g^{}_{S,S}$ (or
  $g^{}_{AS,AS}$) $=p^{}_{1,2}/p^{}_1 p^{}_2$, where $p^{}_{1,2}$ and $\left\{
  p^{}_1, p^{}_2 \right\}$ are the joint and individual probability of
  detecting one photon by the two detectors, respectively}.
\end{thebibliography}

\end{document}